\documentclass[prd,twocolumn,showpacs,nofootinbib]{revtex4-1}
\pdfoutput=1

\usepackage{amsmath}
\usepackage{amsfonts}
\usepackage{wasysym}
\usepackage{graphicx}
\usepackage{comment}

\usepackage[bookmarksopen=true]{hyperref}

\numberwithin{equation}{section}

\def\vect{\boldsymbol}
\def\apgr{\apprge}
\def\aple{\apprle}

\def\filetype{pdf}

\begin{document}

\title{Vortices in Bose-Einstein Condensate Dark Matter}
\author{Ben Kain and Hong Y. Ling}
\affiliation{Department of Physics and Astronomy, Rowan University\\201 Mullica Hill Road, Glassboro, NJ 08028 USA}

\begin{abstract}
\noindent If dark matter in the galactic halo is composed of bosons that form a Bose-Einstein condensate then it is likely that the rotation of the halo will lead to the nucleation of vortices.  After a review of the Gross-Pitaevskii equation, the Thomas-Fermi approximation and vortices in general, we consider vortices in detail.  We find strong bounds for the boson mass, interaction strength, the shape and quantity of vortices in the halo, the critical rotational velocity for the nucleation of vortices and, in the Thomas-Fermi regime, an exact solution for the mass density of a single, axisymmetric vortex.
\end{abstract}


\maketitle


\section{Introduction}

The observational evidence for the existence of non-baryonic dark matter is impressive.  Precision measurements of anisotropies in the cosmic microwave background reveal a universe composed of 4\% baryons, 22\% non-baryonic dark matter and 74\% dark energy \cite{wmap}.  Observation of galaxies and galaxy clusters from rotation curves \cite{rubin}, gravitational lensing \cite{refregier} and X-ray spectra \cite{lewis} has lead to a picture in which galaxies are composed of a luminious galactic disk surrounded by a spherical galactic halo of dark matter which comprises roughly 95\% of the total mass of the galaxy \cite{freese}.

Various explanations for dark matter have been proposed.  Since we are interested in a particle explanation we will mention the two most prominent proposals from particle physics: WIMPs (weakly interacting massive particles) \cite{feng} and axions \cite{pqww}.  WIMPs have a mass around a TeV and interact via the weak force.  Weak interactions automatically lead to the correct relic density \cite{feng}.  A well known WIMP candidate comes from supersymmetry and is known as the lightest supersymmetric partner (LSP).  In many scenarios this is the lightest neutralino, which is a linear combination of the fermionic superpartners of the Higgs scalars and neutral electroweak gauge bosons.  Axions are scalar bosons first proposed to solve the strong CP problem in QCD and may be produced, for example, by an initial misalignment of its vacuum angle in the early universe \cite{pww}.

An intriguing possibility is that if dark matter is composed of bosons (which includes, but is not limited to, the axion \cite{sikyan}) then at sufficiently low temperatures the bosons will form a Bose-Einstein condensate (BEC).  Galactic halos could then be gigantic BECs.  Such an idea, that dark matter may be in the form of a BEC, has been proposed by a number of authors \cite{sin,leekoh,hubak,silmal,wang,mfs} and has been shown to alleviate known problems with the standard cold dark matter picture, such as cuspy central densities and an abundance of small scale structure in halos  \cite{hubak}.  The standard approach to treating a non-relativistic BEC is to use the Gross-Pitaevskii equation, also known as a nonlinear Schr\"ondinger equation, which includes a self-interaction term and a confining, trap potential.  For gravitating BECs the trap potential is simply the Newtonian gravitational potential, so that the Gross-Pitaevskii equation is coupled with the Poisson equation \cite{wang, BH, brook}.

A rotating BEC is known to give rise to a lattice of quantized vortices \cite{expBEC, Fetter}, without which the condensate would be irrotational.  The formation of vortices leads to lower energy states and thus is energetically preferred.  If we are imagining the galactic halo to be made up of bosonic dark matter in the form of a BEC the rotation of the halo may give rise to the formation of vortices \cite{silmal, yumor, brook, shapiro}.

If vortices form in the halo, they may affect the mass distribution and other properties.  In this paper we consider vortices in BEC dark matter in detail. In the next section we review the Gross-Pitaevskii equation, the Thomas-Fermi approximation and vortices.  These equations will be the framework in which all of our work is done.  In section \ref{massbound} we consider the condensate as a whole, populated by a lattice of vortices, and make a model independent argument that for a stable condensate with vortices the boson should have a mass centered around $m\sim10^{-60}-10^{-58} \text{ kg} \sim 10^{-24}-10^{-22} \text{ eV}$.  In section \ref{1vortex} we analyze a single, axisymmetric vortex and, in the Thomas-Fermi regime, find an exact solution for the mass density.  Our results lead to a comprehensive picture for both the condensate and vortices, which we present in section \ref{vorstruc}.  Using our vortex solution, in section \ref{criticalvelocity} we compute the critical rotational velocity of the halo such that vortices would form and find that it is likely to have happened.  We conclude in section \ref{conclusion}.

Although much of this paper is theoretical, it is important that we connect our results to data.  When doing so we will use the following numbers from M31 (the Andromeda galaxy) \cite{M31}, which we take to be not unreasonably characteristic of dark matter halos: A total mass of $M\sim 10^{12} M_\odot \sim 10^{42}$ kg, an average radius of roughly $R \sim 100 \text{ kpc} \sim 10^{21}$ m and an approximate rotational velocity of 125 km/s at 30 kpc.  From this we estimate an average mass density of $\rho_\text{avg}\sim 10^{-23}$ kg/m$^3$ and a rotational angular velocity for the halo of $\Omega \sim 10^{-16}$ rad/s.


\section{Gravitating Bose-Einstein Condensates and Vortices}

In this section we give a general introduction to the Gross-Pitaevskii equation, the Thomas-Fermi approximation and vortices as they apply to gravitating BECs.  We give considerable attention to these equations as all of our results are derived from them.   

\subsection{The Gross-Pitaevskii Equation}

We assume that dark matter is composed of scalar bosons, of mass $m$, that have undergone a phase transition towards Bose-Einstein condensation.  To describe the BEC we use the
standard symmetry-breaking mean-field approach, which is expected to be valid for
systems with a sufficiently large number of particles and at temperatures far
bellow the BEC transition temperature. In this approach one associates an order
parameter, or macroscopic wave function, $\psi(\vect{r})$, with the bosonic field
and ignores high-order correlations due to bosonic quantum field
fluctuations.  The BEC can then be described by the
time-independent Gross-Pitaevskii energy functional:
\begin{equation}
E[\psi]=\int d^{3}r\left(  \frac{\hbar^{2}}{2m}|\nabla\psi|^{2}%
+\frac{1}{2}V_{0}|\psi|^{4}+\frac{1}{2}mV_{G}|\psi|^{2}\right).
\label{gpenergy}
\end{equation}
The first term represents the kinetic energy and the second term 
collisional energy due to repulsive contact interactions
with strength
\begin{equation}
		V_{0}=\frac{4\pi\hbar^{2}a}{m},
\end{equation}
where $a$ $(>0$) is the s-wave scattering length.  The final term is
the gravitational potential energy, with gravitational potential
\begin{equation}
	V_G(\vect{r}) = -Gm\int d^3 r' \frac{|\psi(\vect{r}')|^2}{|\vect{r}-\vect{r}'|},
\end{equation}
which satisfies Poisson's equation:
\begin{equation}
\nabla^{2}V_{G}=4\pi Gm|\psi|^{2}. \label{pois}%
\end{equation}
The condensate wave function, $\psi$, obeys the Gross-Pitaevskii equation
\begin{equation}
\left(  -\frac{\hbar^{2}}{2m}\nabla^{2}+V_{0}|\psi|^{2}+mV_{G}\right)
\psi=\mu\psi, \label{gpeq}%
\end{equation}
which is obtained either by variation of $E[\psi]$ with respect to $\psi
^{\ast}$ under the constraint that the total number of particles, $N$, is conserved (i.e. $\int d^{3}r \, |\psi|^{2}=N$), or equivalently by
variation of the grand-canonical thermodynamic potential, $E[\psi]-\mu N$, with
respect to $\psi^{\ast}$, where $\mu$ is a Lagrange multiplier representing
the chemical potential. \ Our system is then described by the coupled
differential equations (\ref{gpeq}) and (\ref{pois}).

It is very common, and useful, to write the wave function in terms of its modulus and phase:
\begin{equation}
	\psi(\vect{r}) = \left| \psi(\vect{r}) \right| e^{i S(\vect{r})},
\end{equation}
both of which are real, describing a condensate with local velocity
\begin{equation}\label{veldef}
	v(\vect{r}) = \frac{\hbar}{m}\nabla S(\vect{r})
\end{equation}
and mass density $\rho(\vect{r}) = m|\psi(\vect{r})|^2$ with normalization
\begin{equation}
	\int d^3r \, \rho = M,
\end{equation}
where $M$ is the total mass of the condensate.

In terms of these hydrodynamic variables the Gross-Pitaevskii equation (\ref{gpeq}) and the Poisson equation (\ref{pois})
become
\begin{align}
0 &  =\nabla\cdot\left(  \rho\vect{v}\right)  ,\label{gpeq2}\\
0 &  =-\frac{\hbar^{2}}{2m}\frac{\nabla^{2}\sqrt{\rho}}{\sqrt{\rho}}+\frac
{1}{2}m\vect{v}^{2}+\frac{V_{0}}{m}\rho+mV_{G}-\mu,\label{gpeq3}\\
0 &  =\nabla^{2}V_{G}-4\pi G\rho,\label{pois2}%
\end{align}
and the energy functional (\ref{gpenergy}) takes the form
\begin{equation}
E[\rho,\vect{v}]=\int d^{3} r \,\left(  \frac{\hbar^{2}}{2m}\left\vert
\nabla\sqrt{\rho}\right\vert ^{2}+\frac{\vect{v}^{2}}{2}\rho 
+\frac{V_{0}}{2m^{2}}\rho^{2}+\frac{V_{G}}{2}\rho\right)  ,\label{gpenergy1}%
\end{equation}
where (\ref{gpeq2}) and (\ref{gpeq3}) can be obtained either from
equating the imaginary and real parts of the Gross-Pitaevskii equation
(\ref{gpeq}) or by minimizing (\ref{gpenergy1}) with respect to
$\vect{v}$ and $\rho$.

Aside from the first term in (\ref{gpeq3}) (which is called the quantum pressure term), Eqs. (\ref{gpeq2}) and (\ref{gpeq3}) make up the (time-independent)
hydrodynamic description of the condensate since (\ref{gpeq2}) is the
continuity equation and (\ref{gpeq3}) is the Euler equation from
classical fluid dynamics. The quantum pressure term in (\ref{gpeq3}) can
be traced to the first term in (\ref{gpenergy1}) which originates from the
uncertainty principle and hence cannot find its analog in classical physics.  In contrast to the other terms in (\ref{gpenergy1}) this
quantum kinetic energy contains the gradient of the mass density $\rho\left(  \vect{r}\right)  $.  As a result, as the number
of particles in the condensate, or equivalently the size of the wave function,
increases the quantum kinetic energy becomes negligible compared to other
energy contributions except near boundaries of the condensate.  Neglecting the quantum pressure term (i.e. the quantum kinetic energy) is known as the
Thomas-Fermi approximation.  It is important to realize that in our case gravity does not constitute an external trap fixed, a priori, by an external
agent.  Instead it depends on the wave function and hence must be
determined self-consistently with the help of the three coupled equations
(\ref{gpeq2}), (\ref{gpeq3}) and (\ref{pois2}).  In analogy to the ordinary
Thomas-Fermi regime which is defined to be dominated by the balance between
the external trap potential and the repulsive s-wave scattering, we define, as
in \cite{ogka}, the gravitational Thomas-Fermi regime as the regime dominated
by the balance between the gravitational field created by the condensate
particles and the repulsive s-wave scattering.  In this regime the quantum pressure term may be ignored and (\ref{gpeq3}) reduces to
\begin{equation}\label{gpeq4}
	0=\frac{1}{2}m \vect{v}^2  +\frac{V_0}{m}\rho + mV_G - \mu
\end{equation}
and the energy functional (\ref{gpenergy1}) to
\begin{equation} \label{tfenergy}
	E[\rho,\vect{v}] = \int d^3 r \left(\frac{1}{2}\vect{v}^2 + \frac{V_0}{2m^2}\rho + \frac{1}{2}V_G  \right)\rho.
\end{equation}


\subsection{Vortices}
\label{vortices}

To motivate the need for vortices note from (\ref{veldef}) that since the velocity of the condensate is defined in terms of a gradient, its curl vanishes:
\begin{equation} \label{irro}
	\nabla \times \vect{v} = \frac{\hbar}{m} \nabla \times (\nabla S) = 0,
\end{equation}
suggesting that a BEC is irrotational.  Since BECs are well known to rotate $\vect{v}$ must contain a singularity, so as to evade a vanishing curl, generalizing the above expression to
\begin{equation} \label{irro2}
	\nabla \times \vect{v} \sim \delta(\vect{r}-\vect{r}_0).
\end{equation}
This singularity is the vortex core, centered at $\vect{r}_0$, and represents the absence of condensate (i.e. the mass density is zero inside the vortex core).

We imagine the vortex to be cylindrical in shape and adopt cylindrical coordinates ($r_{\bot},\phi,z$).  The wave function takes the form
\begin{equation}\label{vortexansatzk}
	\psi(\vect{r}) = |\psi(r_\bot,z)|e^{ik\phi},
\end{equation}
which we refer to as the vortex ansatz and which describes a $k$-quantized vortex.  The $2\pi$-periodicity of the wave function requires $k$ to be an integer.  From (\ref{veldef}) we find
\begin{equation}\label{vorv}
	\vect{v} = \frac{\hbar k}{mr_\bot}\vect{\hat \phi},
\end{equation}
and thus this is a vortex rotating around the $z$-axis.

The circulation is given by
\begin{equation}\label{circ1}
	\kappa = \oint d\vect{\ell}\cdot\vect{v} = \frac{2\pi\hbar k}{m},
\end{equation}
which may also be evaluated using using Stoke's theorem,
\begin{equation}\label{circ2}
	 \kappa = \frac{2\pi\hbar k}{m} =\int d\vect{A} \cdot \left(\nabla\times\vect{v}\right),
\end{equation}
from which we determine the coefficient in (\ref{irro2}).  Since we are considering the case of rotating around the $z$-axis we find
\begin{equation} \label{irro3}
	\nabla \times \vect{v} = \frac{2\pi\hbar k}{m} \delta(\vect{r}-\vect{r}_0)\vect{\hat z}.
\end{equation}

Vortices with $k\neq 1$ are generally unstable and decay into $k=1$ vortices \cite{Fetter}.  As such, we will confine our attention to $k=1$ vortices.


\section{Bounds on the Boson Mass}
\label{massbound}

Both the energy functional (\ref{gpenergy}) and the Gross-Pitaevskii equation (\ref{gpeq}) were written in a non-rotating frame.  Since we are considering rotating galactic halos, it is natural to work in a frame that is rotating with the same angular velocity as the halo and where the potential energy is time-independent.  Such a transformation (to a frame rotating with constant angular velocity $\vect{\Omega}$) modifies the energy functional by
\begin{equation} \label{energyrot}
	E[\psi] \rightarrow E[\psi] - \int d^3 r \, \psi^* \left(\vect{\Omega}\cdot\vect{\widehat{L}}\right)\psi,
\end{equation}
where $\vect{\widehat{L}}=\vect{r}\times (\hbar/i)\nabla$ is the angular momentum operator, so that (\ref{tfenergy}) becomes
\begin{equation} \label{tfenergyrot2}	
	E[\rho, \vect{v}] = \int d^3 r \left[\frac{1}{2}\vect{v}^2 -\vect{v}\cdot\left(\vect{\Omega}\times \vect{r}\right) + \frac{V_0}{2m^2}\rho + V_G  \right]\rho.
\end{equation}
Without loss of generality we assume the frame to be rotating around the $z$-axis so that the constant angular velocity $\vect{\Omega}$ points in the positive $z$-direction.  We adopt cylindrical coordinates $(r_\bot,\phi,z)$ and define
\begin{equation}
	\vect{v}_\Omega \equiv \vect{\Omega}\times\vect{r} = \Omega r_\bot \vect{\hat \phi}.
\end{equation}
We may rewrite (\ref{tfenergyrot2}) as
\begin{equation}\label{tfenergyrot3}
	E[\psi] = \int d^3 r \left[\frac{1}{2}\left(\vect{v}-\vect{v}_\Omega\right)^2 -\frac{1}{2}\vect{v}^2_\Omega + \frac{V_0}{2m^2}\rho + V_G  \right]\rho,
\end{equation}
showing that the value of $\vect{v}$ which minimizes the energy is precisely $\vect{v}=\vect{v}_\Omega$.  Now,	$\nabla \times \vect{v}_\Omega = 2\vect{\Omega}$, while from (\ref{irro3}) $\nabla\times\vect{v} = (2\pi\hbar k/m)\delta(\vect{r}-\vect{r_0})\vect{\hat z}$.  Thus it is not possible for a condensate to acquire the solid-body rotation $\vect{v}=\vect{v}_\Omega$.  Instead it will nucleate a lattice of uniformly spaced vortices in an attempt to mimic solid-body rotation \cite{Fetter}.  In this sense $\vect{v} \approx \vect{v}_\Omega$.

If the BEC has constant angular velocity $\vect{\Omega}$ and is populated by a lattice of $N_\text{vor}$ vortices then from (\ref{irro3})
\begin{equation} \label{irro4}
	\nabla \times \vect{v} = \frac{2\pi\hbar}{m} \sum_{i=1}^{N_\text{vor}} \delta(\vect{r}-\vect{r}_i)\vect{\hat z},
\end{equation}
where we have assumed all vortices have $k=1$ as mentioned at the end of section \ref{vortices}.  The circulation (\ref{circ1}) gives
\begin{equation}
	\kappa_N = \oint d\vect\ell \cdot \vect{v} = 2\pi\Omega R_\text{BEC}^2 = 2\Omega A_\text{BEC},
\end{equation}
where the path was taken around the entire condensate and thus $R_\text{BEC}$ is the radius of the condensate and $A_\text{BEC}$ its area.  Using Stokes theorem, as was done in (\ref{circ2}), and (\ref{irro4}) we find also for the circulation
\begin{equation}
	\kappa_N = \int d\vect{A} \cdot \left(\nabla\times\vect{v}\right) = N_\text{vor} \frac{2\pi\hbar}{m}.
\end{equation}
Equating our two results we determine the areal density of vortices \cite{Feynman}:
\begin{equation}
	n_\text{vor} = \frac{N_\text{vor}}{A_\text{BEC}} = \frac{m\Omega}{\pi\hbar}.
\end{equation}
The inverse of this quantity is the area per vortex.  Assuming it to be circular with radius $R_\text{vor}$, so that $\pi R_\text{vor}^2 = 1/n_\text{vor}$, we find
\begin{equation}\label{intervorR}
	R_\text{vor} = \sqrt{\frac{\hbar}{m\Omega}}.
\end{equation}  

With this equation we can find a rough lower bound on the boson mass.  The size of the vortex must be smaller than the condensate: $R_\text{vor}\leq R_\text{BEC}$.  Since we are assuming that dark matter in the galactic halo is in the form of a BEC, then with the numbers from the end of the introduction we find
\begin{equation}\label{lowerbound}
	m \apgr \frac{\hbar}{R_\text{BEC}^2 \Omega} \sim 10^{-60} \text{ kg} \sim 10^{-24} \text {eV}.
\end{equation}

This is approximately the mass many authors expect the boson to have for BEC dark matter \cite{sin, hubak, silmal, mfs, brook, shapiro}.  The reason is rather simple and can be understand from a simple qualitative argument \cite{sahni, hubak, jwlee}: For an ultralight dark matter particle with a Compton wave length of astrophysical size, the uncertainty principal can suppress the formation of excess small scale structure seen in the standard cold dark matter scenario.  The masses considered obey
\begin{equation} \label{upperbound}
	m \aple 10^{-58} \text{ kg} \sim 10^{-22} \text{ eV}
\end{equation}
(despite the small mass, BEC dark matter behaves like cold dark matter \cite{hubak, jwlee, bernal}).  In combination with (\ref{lowerbound}) we find that for a galactic halo of BEC dark matter with vortices there is surprisingly little parameter space available for the boson mass.  Our model independent arguments have lead to a boson mass centered around
\begin{equation} \label{bosonmass}
	m \sim 10^{-60} - 10^{-58} \text{ kg} \sim 10^{-24} - 10^{-22} \text{ eV}.
\end{equation}


\section{A Single Axisymmetric Vortex}
\label{1vortex}

In this section we solve for the mass density of a single vortex centered on the axis of rotation.  The condensate, in the Thomas-Fermi regime, is described by the Gross-Pitaevskii equation (\ref{gpeq4}),
\begin{equation}
	0=\frac{1}{2}m \vect{v}^2  +\frac{V_0}{m}\rho + mV_G - \mu,
\end{equation}
and the Poisson equation (\ref{pois2}),
\begin{equation}\label{pois3}
	\nabla^2 V_G = 4\pi G\rho.
\end{equation}
We would like to find a vortex solution to these coupled differential equations.  To do so will use a combination of spherical coordinates $(r,\phi,\theta)$ and cylindrical coordinates $(r_\bot,\phi,z)$ and assume the solution to be of the form of the vortex ansatz (\ref{vortexansatzk}) with $k=1$:
\begin{equation}
	\psi(\vect{r}) = \sqrt{\frac{\rho(r, \theta)}{m}}e^{i\phi},
\end{equation}
where $\rho(r,\theta)=\rho(r_{\bot},z)=m|\psi|^2$ is the mass density.  From (\ref{veldef}) the velocity of the condensate is
\begin{equation}\label{vortexvel}
	\vect{v} = \frac{\hbar}{m r_{\bot}} \vect{\hat \phi}
\end{equation}
and the Gross-Pitaevskii equation becomes
\begin{equation} \label{grosspiteq}
	0 = \frac{\hbar^2}{2mr_{\bot}^2} + \frac{V_0}{m}\rho + mV_G - \mu.
\end{equation}
Taking the Laplacian of this equation and combining it with the Poisson equation (\ref{pois3}) we arrive at
\begin{equation} \label{voreq}
	0 =\nabla^2 \rho + \frac{4\pi Gm^2}{V_0} \rho +\frac{2\hbar^2}{V_0}\frac{1}{r_{\bot}^4}.
\end{equation}
Our system is described by this single equation.

It is convenient to write this equation in terms of dimensionless variables, $\Theta$ and $\bar{\ell}=\{ \bar{r},\bar{r}_{\bot},\bar{z} \}$, defined by
\begin{align}
	\Theta &= \frac{V_0^2}{8\pi Gm^2\hbar^2} \rho = \frac{2\pi\hbar^2 a^2}{G m^4} \rho \label{newvar1} \\
	\bar{\ell} &= \sqrt{\frac{4\pi Gm^2}{V_0}}\ell = \sqrt{\frac{G m^3}{\hbar^2 a}}\ell. \label{newvar2}
\end{align}
(\ref{voreq}) may now be written
\begin{equation} \label{denewvar}
	0=\nabla^2 \Theta + \Theta +\frac{1}{\bar{r}_\bot^4}.
\end{equation}

It is important to stress that while the original Gross-Pitaevskii
equation is nonlinear, the equation for the mass density under the
Thomas-Fermi approximation is linear.  This allows us to write the solution
to (\ref{denewvar}),
\begin{equation}
\Theta=\Theta _{h}(\bar{r},\theta)+\Theta _{p}(\bar{r}_{\bot} ),
\end{equation}
as a superposition between $\Theta _{h}(\bar{r},\theta)$---the solution to the homogeneous part of the differential equation---and $
\Theta _{p}(\bar{r}_{\bot} )$---the particular solution originating from the $1/\bar{r}_{\bot} ^{4}$ term.  The homogeneous part is the Helmholtz equation
with solution
\begin{equation}\label{homsol}
\Theta _{h}(\bar{r},\theta)=\sum_{l=0}^{\infty }\left[ a_{l}j_{l}(\bar{r})+b_{l}y_{l}(\bar{r})\right] P_{l}\left( \cos
\theta \right)  ,
\end{equation}
where $a_l$ and $b_l$ are arbitrary constants, $j_l$ and $y_l$ are spherical Bessel functions and $P_l$ are Legendre polynomials.  The particular solution is 
\begin{equation}\label{comsol}
\begin{split}
\Theta_{p} (\bar{r}_{\bot} )=-\frac{\pi }{\bar{r}_{\bot}^{4}}&\left[ J_{0}(\bar{r}_{\bot}
)G_{2,4}^{3,0}\left( \frac{1}{4}\bar{r}_{\bot} ^{2}\left\vert 
\begin{array}{c}
1/2,3 \\ 
1,1,2,1/2%
\end{array}%
\right. \right) \right. \\
& -Y_{0}(\bar{r}_{\bot} )G_{1,3}^{2,0} \left.\left( \frac{1}{4}\bar{r}_{\bot}
^{2}\left\vert 
\begin{array}{c}
3 \\ 
1,2,1%
\end{array}%
\right. \right) \right] , 
\end{split} 
\end{equation}%
where $J_{0}$ and $Y_{0}$ are Bessel functions of the first and second kind
and 
\begin{equation}
G_{p,q}^{m,n}\left( z\left\vert 
\begin{array}{l}
a_{1},\ldots ,a_{n},a_{n+1},\ldots ,a_{p} \\ 
b_{1},\ldots ,b_{m},b_{m+1},\ldots ,b_{q}%
\end{array}%
\right. \right) 
\end{equation}%
is the Meijer G-function \cite{mathematica}, and is well approximated by 
\begin{equation}
\Theta _{p}(\bar{r}_{\bot})\approx -\frac{1}{4\bar{r}_{\bot} ^{2}}.  \label{mgexp}
\end{equation}

In order to fix the constants in (\ref{homsol}) we note that since the influence of the rotational energy (the $1/\bar{r}_{\bot}
^{4}$ term in (\ref{denewvar})) is vanishingly small in the large $\bar{r}_{\bot} $
limit, the vortex solution must match the vortex free solution for large $
\bar{r}_{\bot} $ in the Thomas-Fermi limit.  This observation translates our task to
finding all constants for the vortex free solution $\Theta
_{h}(\bar{r},\theta)$.  The vortex free solution is well-known in the
literature \cite{chan}, having the form 
\begin{equation}\label{vorfreesol}
\Theta _{h}(\bar{r})=\Theta _{0}\frac{\sin \bar{r}}{\bar{r}},
\end{equation}%
which is obtained from (\ref{homsol}) by requiring it to be spherically symmetric and finite at the origin which sets all constants to zero except $a_{0}\equiv \Theta _{0}$. 

The vortex solution is then 
\begin{equation}\label{apsol}
\Theta \approx \Theta _{0}\frac{\sin \bar{r}}{\bar{r}}-\frac{1}{4\bar{r}_{\bot}^2}. 
\end{equation}

\section{Condensate and Vortex Structure}
\label{vorstruc}

In this section we compile and analyze our results in an attempt to say as much as possible about the structure of the condensate and the vortices which reside in it.  We will form a comprehensive and detailed picture.  While we are using the numbers from the end of the introduction, and thus our results are specific to the Andromeda galaxy, we expect them to hold more generally.

In section \ref{massbound} we found that the boson mass is centered around
\begin{equation} \label{bosonmass2}
	m \sim 10^{-60} - 10^{-58} \text{ kg} \sim 10^{-24} - 10^{-22} \text{ eV}.
\end{equation}
Returning to (\ref{intervorR}) and using these masses, the size of the vortex is around
\begin{equation}
	R_\text{vor} \sim 10^{20} - 10^{21} \text{ m}
\end{equation}
and the number of vortices in the halo is roughly $R_\text{vor}^2/R^2 \sim 1 - 100$, where $R \sim 10^{21}$ m is the radius of the halo from the end of the introduction.

To analyze the shape of the vortex we use the solution (\ref{apsol}).  We expect it be roughly spherical with a thin, cylindrical vortex core along the $z$-axis.  Since the vortex core is along the $z$-axis, to determine its size we analyze our solution (\ref{apsol}) for small $\bar{r}_{\bot}$.  The region of negative mass density is taken to represent the core itself, the mass density being negative because the Thomas-Fermi approximation is invalid in this region.  We will use the standard technique of cutting off the solution, removing the region of negative mass density and setting this region equal to zero.  Thus, the vortex core is defined by the smallest values of $\bar{r}_{\bot \text{c}}$ which satisfy
\begin{equation}
	\Theta_0 \frac{\sin\left(\sqrt{\bar{r}_{\bot \text{c}}^2 + \bar{z}^2}\right)}{\sqrt{\bar{r}_{\bot \text{c}}^2 + \bar{z}^2}} - \frac{1}{4\bar{r}_{\bot \text{c}}^2} = 0.
\end{equation}
The $\bar{z}$-dependence, around $z=0$ and for small $\bar{r}_{\bot \text{c}}$, is
\begin{equation}
	\bar{r}_{\bot \text{c}}(\bar{z}) = \frac{1}{2\sqrt{\Theta}}\sqrt{\frac{\bar{z}}{\sin \bar{z}}}.
\end{equation}
The radius is smallest in the $z=0$ plane:
\begin{equation}\label{coresize}
	\bar{r}_{\bot \text{c}} = \frac{1}{2\sqrt{\Theta_0}} \qquad \text{(for $z=0$)}.
\end{equation}
In the unscaled unit system this equation simply corresponds to the healing length, $R_\text{c}=\sqrt{m/8\pi a \rho_{0}}$, where $\rho_{0}$ is the mass density at the center of a vortex free condensate.

The edge of the vortex is taken to be the first point, outside the core, where the mass density is once again zero \cite{chan,wang,BH}.  Away from the upper and lower caps (in the $z$-direction) the $-1/4\bar{r}_{\bot}^2$ term is negligible and the edge of the vortex is defined by
\begin{equation}
	\Theta_0\frac{\sin \bar{r}_\text{vor}}{\bar{r}_\text{vor}} = 0,
\end{equation} 
with solution $\bar{r}_\text{vor}=\pi$.  Thus, the vortex is roughly spherical in shape with radius $\pi$ or, in the unscaled unit system, 
\begin{equation}
R_{\text{vor}}\ =\pi \sqrt{\frac{\hbar ^{2}a}{Gm^{3}}}.  \label{vortexR}
\end{equation}%

\begin{figure}
	\centering
		\includegraphics{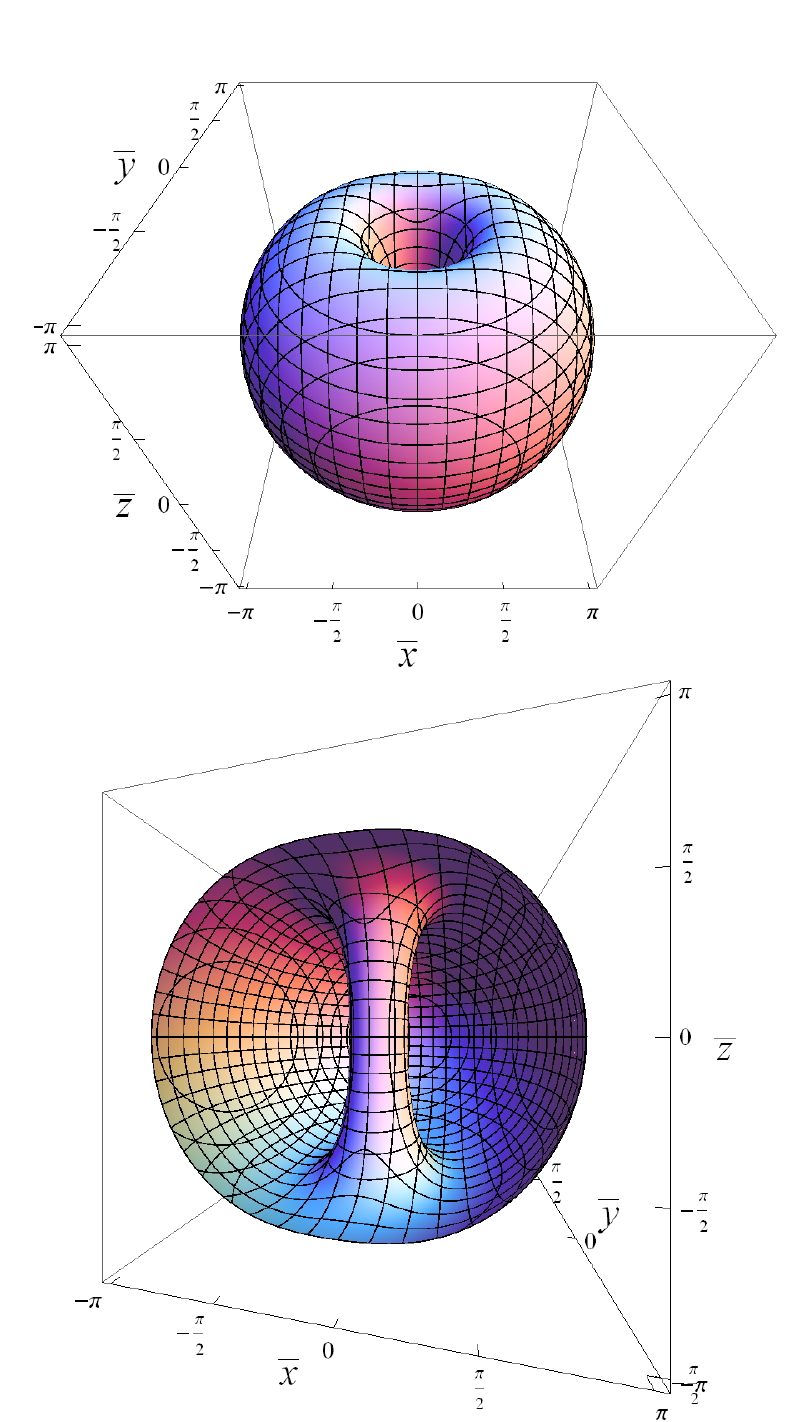}
	\caption{The edge of the vortex and the edge of the vortex core are plotted for $\Theta_0=1$.  The figure on the top shows the spherical shape of the vortex with the opening to the vortex core visible.  The figure on the bottom is a cross section showing the cylindrical shape of the core.  The core can be made arbitrarily thin by increasing $\Theta_0$.}
\label{fig:vortexsol}
\end{figure}In figure \ref{fig:vortexsol} we plotted both the edge of the vortex and the edge of the vortex core for $\Theta_0=1$.  The figure on the top shows the spherical shape of the vortex.  The opening to the vortex core is visible.  The figure on the bottom is a cross section showing the core lying along the $z$-axis and cylindrical in shape.  The core can be made arbitrarily thin by increasing $\Theta_0$.  The figure is in good
agreement with the actual vortex state only in the Thomas-Fermi limit where the healing length is much less than the size of the condensate, or
equivalently $R_\text{c}/R_\text{vor} = \bar{r}_{\bot \text{c}}/\pi = 1/ 2\pi \sqrt{\Theta _{0}} \ll 1$.

We may use (\ref{vortexR}) to determine the characteristic size of the scattering length $a$.  Comparing (\ref{vortexR}) to the characteristic size of vortices given by (\ref{intervorR}) we find that the bosons should obey
\begin{equation}
	a \sim \frac{G}{\hbar \Omega} m^2.
\end{equation}
In lieu of our expected boson mass (\ref{bosonmass2}) the scattering length is expected to be centered around
\begin{equation}
	a \sim 10^{-80} - 10^{-76} \text{ m}.
\end{equation}

Determining the characteristic radius of the vortex core from (\ref{coresize}) requires determining $\Theta_0$.  To determine $\Theta_0$ we might try fitting our solution to data, but since we expect the galactic halo to be populated by many vortices it is not simple to fit the solution for a single vortex to data for an entire galactic halo.  However, we may determine $\Theta_0$ if we make the following assumption: We assume that the average mass density of our vortex solution is characteristic of the average mass density of the galactic halo in which it resides.  In the Thomas-Fermi regime the core has a negligible affect on the total mass and we find
\begin{equation}
	\rho_\text{avg} = \frac{4\pi}{(4/3)\pi R_\text{vor}^3} \int_0^{R_\text{vor}} dr \, r^2 \rho(r) = \frac{3 G m^4}{2\pi^3 \hbar^2 a}\Theta_0.
\end{equation}
From (\ref{coresize}) the vortex core size is then
\begin{equation} \label{rhoavg}
	\bar{r}_{\bot \text{c}} = \frac{1}{2\sqrt{\Theta_0}} = \frac{1}{2} \left(\frac{3Gm^4}{2\pi^3\hbar^2 a^2 \rho_\text{avg}}\right)^{1/2}.
\end{equation}
\begin{figure}
	\centering
		\includegraphics[width=3.3 in]{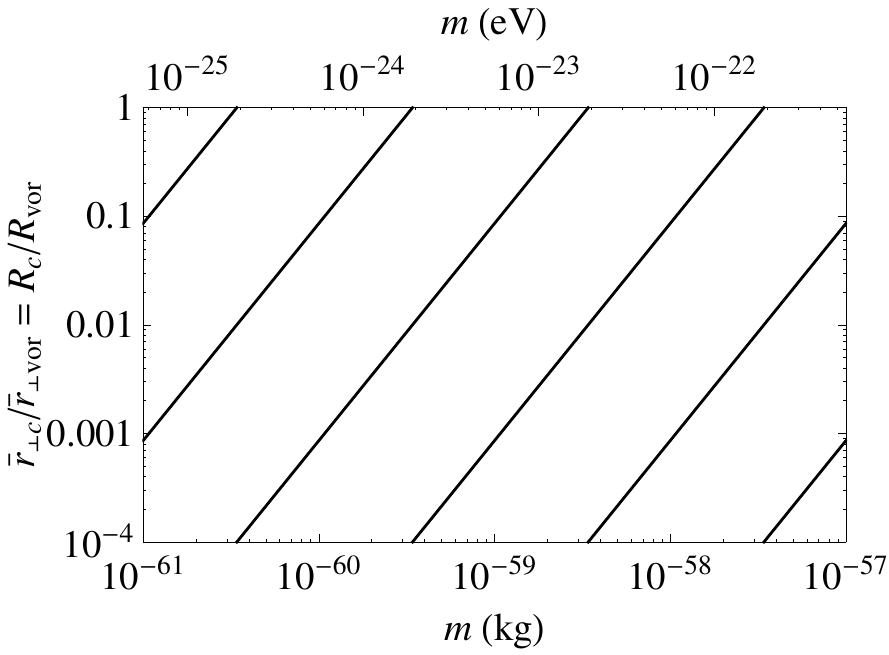}
	\caption{The radius of the vortex core, in the $z=0$ plane, as a fraction of the radius of the vortex is plotted versus the boson mass.  The various lines correspond to different scatting lengths.  From upper left to lower right: $a=10^{-82}$ m, $10^{-80}$ m, $10^{-78}$ m, $10^{-76}$ m, $10^{-74}$ m and $10^{-72}$ m.}
\label{fig:coresize}
\end{figure}In figure \ref{fig:coresize} we have used (\ref{rhoavg}) and, from the introduction, $\rho_\text{avg} \sim 10^{-23}$ kg/m$^3$ to plot the radius of the core in the $z=0$ plane as a fraction of the radius of the vortex versus boson mass.  The different lines in the figure correspond to different values of the scattering length, $a$, as explained in the caption.

Our results have lead to a comprehensive picture of the condensate and vortices.  To reiterate, if dark matter in the galactic halo is composed of bosons in the form of a BEC that has formed vortices then we should expect the boson mass to be centered around $m \sim 10^{-60} - 10^{-58} \text{ kg} \sim 10^{-24} - 10^{-22} \text{ eV}$, the scatting length around $a \sim 10^{-80} - 10^{-76} \text{ m}$, the radius of the vortex around $10^{20} - 10^{21} \text{ m}$, which corresponds to roughly $1-100$ vortices in the halo, and a vortex core radius as displayed in figure \ref{fig:coresize}.  The vortex is expected to have a shape that is nearly spherical with a thin, cylindrical vortex core along the axis of rotation, as displayed in figure \ref{fig:vortexsol}, and a mass density described by (\ref{apsol}).


\section{Critical Velocity for Vortex Formation}
\label{criticalvelocity}

The vortex solution in section \ref{1vortex} was obtained in a non-rotating frame.  The fact that the galactic halo rotates with angular velocity $\vect{\Omega}$ motivates the question of whether a given value of $\vect{\Omega}$ can sustain a single vortex or an array of vortices.  In this section we calculate the critical angular velocity of the condensate at which vortices begin to form.  To understand how \cite{fesv} recall from (\ref{energyrot}) that when moving to a frame rotating with constant angular velocity $\vect{\Omega}$ the energy transforms as
\begin{equation}\label{energyrot2}
	E \rightarrow E'= E- \int d^3 r \, \psi^* \left(\vect{\Omega}\cdot\vect{\widehat{L}}\right)\psi.
\end{equation}
In the rotating (primed) frame we calculate the energy for a condensate with a single ($k=1$) vortex, which we call $E'_1(\Omega)$.  We also calculate the energy for a vortex free condensate (i.e. the $k=0$ configuration), which we call $E'_0$.  In the latter case, since there is no rotation, the energy is independent of $\Omega$ and may be calculated in either frame ($E'_0=E_0$). The difference in energy,
\begin{equation}
	\Delta E'(\Omega) = E'_1(\Omega) - E'_0 = E'_1(\Omega) - E_0,
\end{equation}
is the energy required to form a vortex at angular velocity $\Omega$.  The formation of vortices will occur when it is energetically preferred, i.e. when $\Delta E'(\Omega) \leq 0$.  The critical velocity for vortex formation is then defined as
\begin{equation}
	\Delta E'(\Omega_\text{c}) = 0,
\end{equation}
and vortex formation will occur for $\Omega \geq \Omega_\text{c}$.

For an axisymmetric trap potential, such as the gravitational potential, there is a particularly nice formula.  In this case, if we take the condensate to be rotating around the $z$-axis the wave function is an eigenfunction of the $\widehat{L}_z$ operator and an axisymmetric vortex has total angular momentum
\begin{equation}
	L_z = \int d^3 r \, \psi^*\widehat{L}_z\psi = \hbar\int d^3r \, |\psi|^2 = \frac{M\hbar}{m},
\end{equation}
where $M$ is the total mass of the condensate.  Thus, from (\ref{energyrot2}) we have
\begin{equation}
	E' = E - \Omega\frac{M\hbar}{m}
\end{equation}
and
\begin{equation}
	\Delta E'(\Omega) = E'_1(\Omega) - E'_0(\Omega) = E_1 - E_0 - \Omega\frac{M\hbar}{m}.
\end{equation}
The critical angular velocity is then given by
\begin{equation}
	\Omega_\text{c} = \frac{m}{M\hbar}(E_1 - E_0)
\end{equation}
and can be conveniently calculated entirely in the laboratory (unprimed) frame.

In the following subsections we calculate the necessary energies and approximate the critical angular velocity at which vortices form.


\subsection{Vortex Free Energy}
\label{vorfreeenergy}

A vortex free condensate (i.e. the $k=0$ configuration), has, from (\ref{vortexansatzk}) and (\ref{veldef}), $\psi(\vect{r}) = |\psi(r_{\bot},z)|$ and no rotation, $\vect{v}=0$.  To calculate the energy we use the energy functional (\ref{tfenergy}):
\begin{equation}
	E_0 =  \int d^3 r\, \left(\frac{V_0}{2m^2}\rho + \frac{1}{2}V_G \right)\rho.
\end{equation}
The corresponding Gross-Pitaevskii equation is
\begin{equation} \label{vfgpeq}
	\mu=\frac{V_0}{m}\rho + mV_G.
\end{equation}
While this mass density, $\rho$, differs from the mass density when there is a vortex it must satisfy the same normalization:
\begin{equation}\label{norm}
	\int d^3r \, \rho = M,
\end{equation}
where $M$ is the mass of the condensate, which doesn't change as a vortex is formed.

Multiplying (\ref{vfgpeq}) by $\rho$ and integrating we find
\begin{equation}
	\mu = \frac{2m}{M}E_0.
\end{equation}
We assume that $\mu$ is constant and thus the right hand side of (\ref{vfgpeq}) must be independent of position.  This means we are at liberty to evaluate the right hand side at any point inside the condensate we'd like.  We choose to evaluate it in the $z=0$ plane at the edge of the condensate, $r_{\bot}=R_\text{BEC}$, where the mass density is zero.  Combining the two equations for $\mu$ we have
\begin{equation}\label{E0con}
	E_0 = \frac{1}{2}M V_G(R_\text{BEC}).
\end{equation}
We will comment on this term below after looking at the contribution from the vortex.


\subsection{Single Vortex Energy}

For a single, axisymmetric vortex we use the vortex ansatz (\ref{vortexansatzk}), with $k=1$, which has velocity squared $\vect{v}^2=\hbar^2/m^2 r_{\bot}^2$.  The energy is still obtained from (\ref{tfenergy}),
\begin{equation} \label{1venergy}
	E_1 = \int d^3 r \, \left(\frac{1}{2}\vect{v}^2 + \frac{V_0}{2m^2}\rho + \frac{1}{2}V_G  \right)\rho,
\end{equation}
and the Gross-Pitaevskii equation is
\begin{equation} \label{1vgpeq}
	\mu=\frac{1}{2}m\vect{v}^2 +\frac{V_0}{m}\rho + mV_G .
\end{equation}
While this mass density, $\rho$, differs from the mass density for the vortex free condensate considered above, it must satisfy the same normalization (\ref{norm}) with the same mass, $M$.

Multiplying (\ref{1vgpeq}) by $\rho$ and integrating we find
\begin{equation}
	\mu = \frac{m}{M}\left[2E_1 - \int d^3 r\, \left(\frac{1}{2}\vect{v}^2\right)\rho\right].
\end{equation}
As before the right hand side of (\ref{1vgpeq}) must be independent of position and we choose to evaluate it in the $z=0$ plane at the edge of the vortex, $r_{\bot}=R_\text{vor}$, where the mass density is zero.  Combining the two equations for $\mu$ we have
\begin{equation}
	E_1 = \frac{1}{2} M V_G (R_\text{vor}) + \frac{1}{4} M \vect{v}^2 (R_\text{vor}) + E_{1,\vect{v}},
\end{equation}
where
\begin{equation}
	E_{1,\vect{v}} =  \frac{1}{2} \int d^3 r \, \left(\frac{1}{2}\vect{v}^2\right) \rho.
\end{equation}

$E_{1,\vect{v}}$ appears to be impossible to evaluate analytically.  Building upon the intuition gained from figure \ref{fig:vortexsol} we estimate the integral by approximating the vortex solution (\ref{apsol}) with the vortex free solution (\ref{vorfreesol}) and an empty cylindrical core along the $z$-axis with constant radius $\bar{r}_{\bot \text{c}}$.  This estimation gives an upper bound on the integral.  Under this approximation $E_{1,\vect{v}}$ amounts to, in spherical coordinates,
\begin{widetext}
\begin{equation}
	E_{1,\vect{v}} = \frac{\hbar}{8\pi}\sqrt{\frac{Gm}{a^3}} \int d\bar{r} d\theta d\phi \, \bar{r}^2\sin\theta \frac{\Theta}{\bar{r}_{\bot}^2} 
	=
	\frac{\hbar}{8\pi}\sqrt{\frac{Gm}{a^3}} \Theta_0 \int\limits_0^{2\pi} d\phi \int\limits_{\tan^{-1}(\bar{r}_{\bot \text{c}}/\pi)}^{\pi-\tan^{-1}(\bar{r}_{\bot \text{c}}/\pi)} \frac{d\theta}{\sin\theta}
	\int\limits_{\bar{r}_{\bot \text{c}}/\sin\theta}^{\pi} d\bar{r}\, \frac{\sin \bar{r}}{\bar{r}}.
\end{equation}
\end{widetext}
To leading order in $\bar{r}_{\bot \text{c}}$ this evaluates to
\begin{equation}
	E_{1,\vect{v}} \approx \frac{M\hbar}{m} \Omega_{\text{c},0} \left[\text{Si}(\pi) \ln \left(2\frac{\pi}{\bar{r}_{\bot \text{c}}}\right) +\Gamma \right],
\end{equation}
where $\text{Si}(\pi) \approx 1.852$ is the sine integral evaluated at $\pi$,
\begin{equation}\label{gamma}
	\Gamma = -\pi + \frac{\pi^3}{3^2\cdot3!} - \frac{\pi^5}{5^2\cdot 5!} +\cdots \approx -2.658,
\end{equation}
and
\begin{equation}\label{omegac0}
	\Omega_{\text{c},0} = \frac{1}{8 M}\sqrt{\frac{Gm^3}{a^3}} \frac{1}{\bar{r}_{\bot \text{c}}^2}= \frac{\pi}{4}\frac{\hbar}{m R_\text{BEC}^2},
\end{equation}
where to obtain the second equality we used (\ref{coresize}), (\ref{vortexR}), (\ref{rhoavg}) and $\rho_\text{avg} = M/(4/3)\pi R_\text{vor}^3$ and replaced $R_\text{vor}$ with $R_\text{BEC}$ since we are assuming that a single vortex is forming over the entire condensate.

Putting everything together:
\begin{equation} \label{E1con}
\begin{split}
	E_1 \approx  &\frac{1}{2} M V_G (R_\text{vor}) \\
	&+\frac{1}{\pi}\frac{M\hbar}{m} \Omega_{\text{c},0} + \frac{M\hbar}{m} \Omega_{\text{c},0} \left[\text{Si}(\pi) \ln \left(2\frac{\pi}{\bar{r}_{\bot \text{c}}}\right) +\Gamma \right].
\end{split}
\end{equation}
It is not difficult to show that the first term, like its counterpart in (\ref{E0con}), is roughly $\hbar\sqrt{Gm/a^3}$, without any factors of $\bar{r}_{\bot \text{c}}$ in the denominator, and thus negligible.  The second line above is then the dominant contribution.


\subsection{Critical Angular Velocity}

The critical angular velocity for the formation of vortices is
\begin{equation}\label{criticalvel}
	\Omega_\text{c} = \frac{m}{M\hbar}(E_1 - E_0) \approx \Omega_{\text{c},0} \left[\text{Si}(\pi) \ln \left(2\frac{\pi}{\bar{r}_{\bot \text{c}}}\right) +\frac{1}{\pi} + \Gamma \right],
\end{equation}
where from (\ref{rhoavg})
\begin{equation}
	\bar{r}_{\bot \text{c}} =\frac{1}{2} \left(\frac{3Gm^4}{2\pi^3\hbar^2 a^2 \rho_\text{avg}}\right)^{1/2}
\end{equation}
and $\Omega_{\text{c},0}$ is given in (\ref{omegac0}) and $\Gamma$ in (\ref{gamma}).
\begin{figure}
	\centering
		\includegraphics[width=3.1 in]{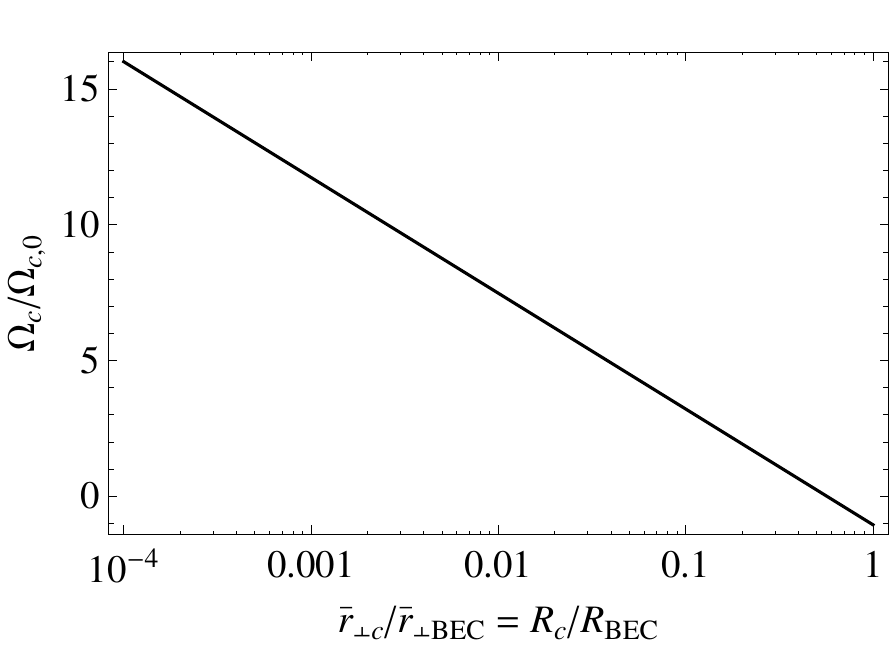}
	\caption{$\Omega_\text{c}/\Omega_{\text{c},0}$ is plotted versus the radius of the core in the $z=0$ plane as a fraction of the radius of the condensate.}
\label{fig:critvel}
\end{figure} In figure \ref{fig:critvel} we plotted $\Omega_\text{c}/\Omega_{\text{c},0}$ versus the radius of the core in the $z=0$ plane as a fraction of the radius of the condensate.  One can see that $\Omega_\text{c}/\Omega_{\text{c},0} \aple O(10)$.  With the numbers from the end of the introduction and our range of boson masses (\ref{bosonmass2}) we have $\Omega_{\text{c},0} \sim 10^{-18} - 10^{-16}$ rad/s and thus there exists large portions of parameter space where the Andromeda galaxy would have formed vortices.


\section{Conclusion}
\label{conclusion}
We considered the possibility that dark matter in a galactic halo is composed of bosons in the form of a Bose-Einstein condensate.  Rotation of the halo is likely to nucleate vortices and we studied such vortices in detail.  We found, in the Thomas-Fermi regime, an exact solution for the mass density of a single, axisymmetric vortex and, by fitting our equations to astrophysical data from the Andromeda galaxy, developed a comprehensive picture for the condensate and vortices and strong bounds for the parameters.  If dark matter in the galactic halo is composed of bosons in the form of a BEC with vortices then model independent arguments require a boson mass centered around $m \sim 10^{-60} - 10^{-58} \text{ kg} \sim 10^{-24} - 10^{-22} \text{ eV}$ and a radius around $10^{20} - 10^{21} \text{ m}$, which corresponds to roughly $1-100$ vortices in the halo.  Combining this with our solution, the scatting length is centered around $a \sim 10^{-80} - 10^{-76} \text{ m}$.  Using our solution we also determined the size of the vortex core and calculated the critical rotational velocity for the nucleation of vortices, showing that there exists large portions of parameter space where vortices would form.  While these results are specific to the Andromeda galaxy, since the Andromeda galaxy is not atypical we expect them to hold more generally.


\section*{Acknowledgments}

We are grateful to Eddie Guerra for helpful discussions.  H. Y. L. was supported in part by the US National Science Foundation and US Army Research Office.


\end{document}